\documentclass{article}

\begin{document}

\centerline{{\LARGE No-deleting principle for two unitary copies }}

\centerline{Dafa Li}
\centerline{Department of Mathematical Sciences, Tsinghua
University, Beijing 100084, CHINA} \centerline{Corresponding author: Dafa Li}
\centerline{email: lidafa@tsinghua.edu.cn}

Abstract: Pati and Braunstein defined a deleting machine and showed the
impossibility of deleting one of two identical copies of an unknown quantum
state. So far, no one has defined two non-identical copies of a quantum
state, of course no one has discussed the impossibility of deleting one of
two non-identical copies of an unknown quantum state. In this paper, we
define $\mathcal{U}|\psi \rangle $ and $U|\psi \rangle $, where $\mathcal{U}$
and $U$\ are any unitary operators, as two unitary copies of a quantum state
$|\psi \rangle $, and show that it is impossible to delete one of two
unitary copies of an unknown state.

\section{Introduction}

Quantum information theory has revealed several fundamental differences
between classical and quantum systems. In classical information theory,
information encoded in strings of bits can be copied and deleted perfectly
without any restriction. However, the laws of quantum mechanics impose
strict limitations on the manipulation of quantum information. One of the
earliest and most important discoveries in this direction was the quantum
no-cloning theorem, proposed by William K. Wootters and Wojciech H. Zurek in
1982 \cite{Wootters}. The theorem states that it is impossible to create an
identical copy of an unknown quantum state. This principle plays a crucial
role in the development of quantum computing, quantum teleportation, quantum
cryptography, and quantum key distribution.

The cloning machine was defined as follows \cite{Wootters}.

\begin{equation}
T|\psi \rangle |0\rangle =|\psi \rangle |\psi \rangle .  \label{cp-1}
\end{equation}%
\ \ \ By using a unitary evolution $T$, the aim of the cloning machine is to
make an identical copy of an unknown quantum state.

A copy machine at an office with color, black-and-white, and scaling
capabilities allows users to adjust document sizes and colors via the
control panel. In this paper,\ we call $\mathcal{U}|\psi \rangle $ the
unitary copy of the state $|\psi \rangle $, where $\mathcal{U}$ is any
unitary operator. In \cite{DLi}, we defined the following cloning machine
with unitary scaling as follows and showed that it is impossible to make a
unitary copy of an unknown quantum state by using a unitary evolution.

\begin{equation}
T|\psi \rangle |0\rangle =|\psi \rangle \mathcal{U}|\psi \rangle
\label{cp-2}
\end{equation}

The impossibility of cloning naturally leads to another important question:
can an unknown quantum state be deleted? This question gave rise to the
quantum no-deleting theorem, which states that there exists no quantum
deleting machine capable of deleting one copy of an unknown quantum state
against another identical copy \cite{Pati}.

In 2000, Pati and Braunstein defined the following quantum deleting machine
\cite{Pati}.
\begin{equation}
|\psi \rangle _{A}|\psi \rangle _{B}|A\rangle _{C}\rightarrow |\psi \rangle
_{A}|\Sigma \rangle _{B}|A_{\psi }\rangle _{C}  \label{pb}
\end{equation}%
The deleting machine in Eq. (\ref{pb}) is called the standard deleting
machine. Pati and Braunstein showed that the deleting machine cannot delete
one of two identical copies of an unknown quantum state.

In \cite{Qiu}, Qiu discussed Some analogies between quantum cloning and
quantum deleting. No-hiding theorem and no-broadcasting were proposed in
\cite{Samuel} and \cite{Barnum}, respectively.

\section{Deleting machine for two unitary copies}

In \cite{Pati}, Pati and Braunstein indicated that the standard deleting
machine is not defined for two non-identical input states. So far, no one
has defined two non-identical copies of a quantum state. In this paper,\ we
call $\mathcal{U}|\psi \rangle $ and $U|\psi \rangle $ the unitary copies of
the state $|\psi \rangle $ for any unitary operators $\mathcal{U}$ and $U$.

In \cite{Pati}, Pati and Braunstein defined the deleting machine for two
identical copies of an unknown state. We propose the following deleting
machine for two unitary copies $\mathcal{U}|\psi \rangle $ and $U|\psi
\rangle $ of an unknown state $|\psi \rangle $, where $\mathcal{U}$ \ and $U$
are any unitary operators, such that the linear operator\ acts on the
combined Hilbert space of input and ancilla.

\begin{equation}
T(\mathcal{U}|\psi \rangle _{A}U|\psi \rangle _{B}|A\rangle _{C})=\mathcal{U}%
|\psi \rangle _{A}|\Sigma \rangle _{B}U|A_{\psi }\rangle _{C}  \label{u-1}
\end{equation}%
Adopting the notations in \cite{Pati}, $|A\rangle _{C}$ is the initial
ancilla, $|\Sigma \rangle _{B}$ is a standard state, and $U|A_{\psi }\rangle
_{C}$ is the final state of the ancilla, where $|A_{\psi }\rangle _{C}$ may
depend on the state $|\psi \rangle $. The deleting machine in Eq. (\ref{u-1}%
) is called the deleting machine for two unitary copies.

The aim of the deleting machine is to delete one of two unitary copies of an
unknown state and replace it with some standard state of a qubit $|\Sigma
\rangle $. Here, it excludes swapping the second and the third states as
mentioned in \cite{Pati, Qiu}. When $U=$ $\mathcal{U=I}$, where $I$ is the
identity,\ the deleting machine for two unitary copies in Eq. (\ref{u-1})
reduces to the standard deleting machine in Eq. (\ref{pb}). We show that the
deleting machine for two unitary copies in Eq. (\ref{u-1}) cannot delete one
of two unitary copies of an unknown state\ below.\

Let $|H\rangle $ and $|V\rangle $\ be a pair of horizontally or vertically
polarized photons. Then, from Eq. (\ref{u-1}), obtain
\begin{eqnarray}
T(\mathcal{U}|H\rangle _{A}U|H\rangle _{B}|A\rangle _{C}) &=&\mathcal{U}%
|H\rangle _{A}|\Sigma \rangle _{B}U|A_{H}\rangle _{C}  \label{u-2} \\
T(\mathcal{U}|V\rangle _{A}U|V\rangle _{B}|A\rangle _{C}) &=&\mathcal{U}%
|V\rangle _{A}|\Sigma \rangle _{B}U|A_{V}\rangle _{C}  \label{u-3}
\end{eqnarray}%
Let $|\psi \rangle $ be an arbitrary unknown state:
\begin{equation}
|\psi \rangle =\alpha |H\rangle +\beta |V\rangle ,
\end{equation}%
where $\alpha $ and $\beta $ are arbitrary complex numbers such that $%
\left\vert \alpha \right\vert ^{2}+\left\vert \beta \right\vert ^{2}=1$. A
calculation yields
\begin{eqnarray}
&&\mathcal{U}|\psi \rangle _{A}U|\psi \rangle _{B}|A\rangle _{C}  \nonumber
\\
&=&\alpha ^{2}\mathcal{U}|H\rangle _{A}U|H\rangle _{B}|A\rangle _{C}+\beta
^{2}\mathcal{U}|V\rangle _{A}U|V\rangle _{B}|A\rangle _{C}  \nonumber \\
&&+\alpha \beta (\mathcal{U}|H\rangle _{A}U|V\rangle _{B}+\mathcal{U}%
|V\rangle _{A}U|H\rangle _{B})|A\rangle _{C}.  \label{u-5}
\end{eqnarray}%
Then, via Eqs. (\ref{u-2}, \ref{u-3}, \ref{u-5}) and by the linearity of the
deleting machine, LHS of Eq. (\ref{u-1}) becomes
\begin{eqnarray}
&&T(\mathcal{U}|\psi \rangle _{A}U|\psi \rangle _{B}|A\rangle _{C})
\nonumber \\
&=&\alpha ^{2}\mathcal{U}|H\rangle _{A}|\Sigma \rangle _{B}U|A_{H}\rangle
_{C}+\beta ^{2}\mathcal{U}|V\rangle _{A}|\Sigma \rangle _{B}U|A_{V}\rangle
_{C}+\alpha \beta |\Omega \rangle _{ABC},  \label{u-6}
\end{eqnarray}%
where
\begin{equation}
|\Omega \rangle _{ABC}=T(\mathcal{U}|H\rangle _{A}U|V\rangle _{B}|A\rangle
_{C})+T(\mathcal{U}|V\rangle _{A}U|H\rangle _{B}|A\rangle _{C}).  \label{ome}
\end{equation}%
Here, $|\Omega \rangle _{ABC}$ cannot be determined by the definition in Eq.
(\ref{u-1}) now. But, we can demonstrate that $|\Omega \rangle _{ABC}$ is
uniquely determined by Eqs. (\ref{u-2}, \ref{u-3}) and the linearity of the
deleting machine below. Anyway, one can see that $|\Omega \rangle _{ABC}$ is
independent of $\alpha $ and $\beta $. Therefore, Eq. (\ref{u-6}) is a
homogeneous quadratic polynomial in $\alpha $ and $\beta $.

For $|\psi \rangle =\alpha |H\rangle +\beta |V\rangle $, RHS of Eq. (\ref%
{u-1}) becomes
\begin{equation}
\mathcal{U}|\psi \rangle _{A}|\Sigma \rangle _{B}U|A_{\psi }\rangle
_{C}=(\alpha \mathcal{U}|H\rangle +\beta \mathcal{U}|V\rangle )_{A}|\Sigma
\rangle _{B}U|A_{\psi }\rangle _{C}.  \label{u-7}
\end{equation}%
From the fact that both RHS of Eqs. (\ref{u-6}, \ref{u-7}) should be equal,
RHS of Eq. (\ref{u-7}) must also be a homogeneous quadratic polynomial in $%
\alpha $ and $\beta $. To this end, $|A_{\psi }\rangle _{C}$ in Eq. (\ref%
{u-7})\ must be of the following form,%
\begin{equation}
|A_{\psi }\rangle _{C}=\alpha |A_{H}\rangle _{C}+\beta |A_{V}\rangle _{C}.
\label{u-8}
\end{equation}%
Then, via Eq. (\ref{u-8}), RHS of Eq. (\ref{u-7}) becomes the following:
\begin{eqnarray}
&&(\alpha \mathcal{U}|H\rangle +\beta \mathcal{U}|V\rangle )_{A}|\Sigma
\rangle _{B}U(\alpha |A_{H}\rangle _{C}+\beta |A_{V}\rangle _{C})  \nonumber
\\
&=&\alpha ^{2}\mathcal{U}|H\rangle _{A}|\Sigma \rangle _{B}U|A_{H}\rangle
_{C}+\beta ^{2}\mathcal{U}|V\rangle _{A}|\Sigma \rangle _{B}U|A_{V}\rangle
_{C}  \nonumber \\
&&+\alpha \beta (\mathcal{U}|H\rangle _{A}|\Sigma \rangle _{B}U|A_{V}\rangle
_{C}+\mathcal{U}|V\rangle _{A}|\Sigma \rangle _{B}U|A_{H}\rangle _{C}).
\label{u-9}
\end{eqnarray}

\ From Eqs. (\ref{u-6}, \ref{u-9}), we obtain
\begin{equation}
|\Omega \rangle _{ABC}=\mathcal{U}|H\rangle _{A}|\Sigma \rangle
_{B}U|A_{V}\rangle _{C}+\mathcal{U}|V\rangle _{A}|\Sigma \rangle
_{B}U|A_{H}\rangle _{C}  \label{u-10}
\end{equation}%
and
\begin{eqnarray}
&&T(\mathcal{U}|\psi \rangle _{A}U|\psi \rangle _{B}|A\rangle _{C})
\nonumber \\
&=&\mathcal{U}(\alpha |H\rangle +\beta |V\rangle )_{A}|\Sigma \rangle
_{B}U(\alpha |A_{H}\rangle _{C}+\beta |A_{V}\rangle _{C})  \label{u-11}
\end{eqnarray}%
From \cite{Pati, Qiu},\ RHS of Eq. (\ref{u-11}) must be normalized for all
possible $\alpha $ and $\beta $. Then, $|A_{H}\rangle _{C}$ and $%
|A_{V}\rangle _{C}$ are orthogonal and normal (i.e. $|A_{H}\rangle _{C}$ and
$|A_{V}\rangle _{C}$ are orthonormal).

One can see that the final state of the ancilla is
\begin{equation}
U|A_{\psi }\rangle _{C}=U(\alpha |A_{H}\rangle _{C}+\beta |A_{V}\rangle
_{C}).  \label{u-12}
\end{equation}%
Thus, $U|\psi \rangle _{B}$ is not deleted whereas $U|\psi \rangle _{B}$ is
moved to the ancilla. Here, we show that the deleting machine \ for two
unitary copies\ in Eq. (\ref{u-1}) cannot delete the unitary copy $U|\psi
\rangle $ of an unknown state $|\psi \rangle $.

\section{Via the standard deleting machine define the deleting machine for
two unitary copies and vice versa}

\subsection{Via the standard deleting machine define the deleting machine
for two unitary copies}

Let $U^{H}$ be the Hermitian transpose of $U$. Then, it is easy to see that
\begin{equation}
(\mathcal{U}^{H}\otimes U^{H}\otimes I)(\mathcal{U}|\psi \rangle _{A}U|\psi
\rangle _{B}|A\rangle _{C})=|\psi \rangle _{A}|\psi \rangle _{B}|A\rangle
_{C}
\end{equation}%
Let $T_{PB}$ stand for the standard deleting machine. Then,
\begin{equation}
T_{PB}|\psi \rangle _{A}|\psi \rangle _{B}|A\rangle _{C}=|\psi \rangle
_{A}|\Sigma \rangle _{B}|A_{\psi }\rangle _{C},  \label{pb-2}
\end{equation}%
where $|A_{\psi }\rangle _{C}$ may depend on the state $|\psi \rangle $ \cite%
{Pati}. Let $T_{PB}^{\prime }=T_{PB}(\mathcal{U}^{H}\otimes U^{H}\otimes I)$%
. Then,
\begin{equation}
T_{PB}^{\prime }(\mathcal{U}|\psi \rangle _{A}U|\psi \rangle _{B}|A\rangle
_{C})=|\psi \rangle _{A}|\Sigma \rangle _{B}|A_{\psi }\rangle _{C}
\label{pb-3}
\end{equation}%
Clearly, Eq. (\ref{pb-3}) is not the deleting machine for two unitary
copies\ in Eq. (\ref{u-1}).

Let
\begin{equation}
T_{PB}^{\prime \prime }=(\mathcal{U}\otimes I\otimes U)T_{PB}(\mathcal{U}%
^{H}\otimes U^{H}\otimes I).  \label{pb-4}
\end{equation}%
Then,
\begin{equation}
T_{PB}^{\prime \prime }(\mathcal{U}|\psi \rangle _{A}U|\psi \rangle
_{B}|A\rangle _{C})=\mathcal{U}|\psi \rangle _{A}|\Sigma \rangle
_{B}U|A_{\psi }\rangle _{C}  \label{pb-5}
\end{equation}%
One can see that $T_{PB}^{\prime \prime }$ moves the unitary copy $U|\psi
\rangle _{B}$ of an unknown state $|\psi \rangle $ to the ancilla without
deleting the unitary copy $U|\psi \rangle _{B}$.

\subsection{Via the deleting machine for two unitary copies define the
standard deleting machine}

One can see that
\begin{equation}
(\mathcal{U}\otimes U\otimes I)|\psi \rangle _{A}|\psi \rangle _{B}|A\rangle
_{C}=\mathcal{U}|\psi \rangle _{A}U|\psi \rangle _{B}|A\rangle _{C}
\label{g-1}
\end{equation}%
Then, by using the deleting machine for two unitary copies in Eq. (\ref{u-1}%
), obtain
\begin{eqnarray}
&&T(\mathcal{U}\otimes U\otimes I)|\psi \rangle _{A}|\psi \rangle
_{B}|A\rangle _{C}  \nonumber \\
&=&T(\mathcal{U}|\psi \rangle _{A}U|\psi \rangle _{B}|A\rangle _{C})
\nonumber \\
&=&\mathcal{U}|\psi \rangle _{A}|\Sigma \rangle _{B}U|A_{\psi }\rangle _{C}.
\label{g-3}
\end{eqnarray}
Let $T^{\ast }=T(\mathcal{U}\otimes U\otimes I)$. Then,
\begin{equation}
T^{\ast }|\psi \rangle _{A}|\psi \rangle _{B}|A\rangle _{C}=\mathcal{U}|\psi
\rangle _{A}|\Sigma \rangle _{B}U|A_{\psi }\rangle _{C}  \label{g-4}
\end{equation}%
Clearly, Eq. (\ref{g-4}) is not the standard deleting machine in Eq. (\ref%
{pb}).

Let
\begin{equation}
T^{\ast \ast }=(\mathcal{U}^{H}\otimes I\otimes U^{H})T(\mathcal{U}\otimes
U\otimes I)  \label{g-5}
\end{equation}%
Then,
\begin{equation}
T^{\ast \ast }|\psi \rangle _{A}|\psi \rangle _{B}|A\rangle _{C}=|\psi
\rangle _{A}|\Sigma \rangle _{B}|A_{\psi }\rangle _{C}  \label{g-6}
\end{equation}%
One can know that $T^{\ast \ast }$ in Eq. (\ref{g-6}) moves an identical
copy $|\psi \rangle _{B}$ of an unknown state $|\psi \rangle $ to the
ancilla without deleting the copy $|\psi \rangle _{B}$.

Acknowledgement

Thank Prof. S. Braunstein for helpful comments.


\begin{thebibliography}{9}
\bibitem{Wootters} W. K. Wootters and W. H. Zurek., A Single quantum cannot
be cloned. Nature 299, 802-803 (1982).

\bibitem{Pati} A.K. Pati and S.L. Braunstein, Impossibility of deleting an
unknown quantum state.\ Nature 404, 164 (2000). arxiv, e-print 9911090v2.

\bibitem{DLi} Dafa Li, No-cloning with unitary scaling. Arxiv: e-print
2604.2296.

\bibitem{Qiu} D. Qiu, Some analogies between quantum cloning and quantum
deleting, Phys. Rev. A 65, 052303 (2002)

\bibitem{Samuel} Braunstein, Samuel L. and Pati, Arun K., Quantum
Information Cannot Be Completely Hidden in Correlations: Implications for
the Black-Hole Information Paradox. Phys. Rev. Lett. 98 (8), 080502 (2007).

\bibitem{Barnum} Barnum, Howard; Caves, Carlton M.; Fuchs, Christopher A.;
Jozsa, Richard; Schumacher, Benjamin. \ Noncommuting Mixed States Cannot Be
Broadcast. Phys. Rev. Lett..76 (15), 2818--2821 (1996).
arXiv:quant-ph/9511010.
\end{thebibliography}
\end{document}